\documentclass[aps,prl,reprint,10pt,twocolumn]{revtex4-1}
\usepackage{graphicx}
\usepackage{amssymb}
\usepackage{amsmath}
\usepackage{amsfonts}
\usepackage{color}
\usepackage[latin1]{inputenc}
\usepackage{float}
\usepackage{comment}

\begin{document}
\title{Circular Kardar-Parisi-Zhang interfaces evolving out of the plane}

\author{I. S. S. Carrasco$^{(a,b)}$}
\email{ismael.carrasco@ufv.br}
\author{T. J. Oliveira$^{(a)}$}
\email{tiago@ufv.br}
\affiliation{$(a)$ Departamento de F\'isica, Universidade Federal de Vi\c cosa, 36570-900, Vi\c cosa, Minas Gerais, Brazil \\
$(b)$ Instituto de F\'isica, Universidade Federal Fluminense, 24210-340, Niter\'oi, Rio de Janeiro, Brazil}

\begin{abstract}
Circular KPZ interfaces spreading radially in the plane have GUE Tracy-Widom (TW) height distribution (HD) and Airy$_2$ spatial covariance, but what are their statistics if they evolve on the surface of a different background space, such as a bowl, a cup, or any surface of revolution? To give an answer to this, we report here extensive numerical analyses of several one-dimensional KPZ models on substrates whose size enlarges as $\langle L(t) \rangle = L_0+\omega t^{\gamma}$, while their mean height $\langle h \rangle$ increases as usual [$\langle h \rangle\sim t$]. We show that the competition between the $L$ enlargement and the correlation length ($\xi \simeq c t^{1/z}$) plays a key role in the asymptotic statistics of the interfaces. While systems with $\gamma>1/z$ have HDs given by GUE and the interface width increasing as $w \sim t^{\beta}$, for $\gamma<1/z$ the HDs are Gaussian, in a correlated regime where $w \sim t^{\alpha \gamma}$. For the special case $\gamma=1/z$, a continuous class of distributions exists, which interpolate between Gaussian (for small $\omega/c$) and GUE (for $\omega/c \gg 1$). Interestingly, the HD seems to agree with the Gaussian symplectic ensemble (GSE) TW distribution for $\omega/c \approx 10$. Despite the GUE HDs for $\gamma>1/z$, the spatial covariances present a strong dependence on the parameters $\omega$ and $\gamma$, agreeing with Airy$_2$ only for $\omega \gg 1$, for a given $\gamma$, or when $\gamma=1$, for a fixed $\omega$. These results considerably generalize our knowledge on the 1D KPZ systems, unveiling the importance of the background space in their statistics.
\end{abstract}

\maketitle

In recent years, a renewed interest on interface dynamics has brought due to the interesting observation that some of the related universality classes split into subclasses depending on the geometry of the interfaces. Even though such geometry dependence has been numerically observed for the two-dimensional (2D) Kardar-Parisi-Zhang (KPZ) \cite{KPZ} class \cite{healy12,alves13,healy13,Ismael14}, as well as in the class of the nonlinear molecular beam epitaxy equation \cite{Villain,LDS} in both 1D and 2D \cite{Ismael16}, the main efforts have been concentrated on 1D KPZ systems (see \cite{Corwin-RMTA2012,Takerev} for recent reviews), which will be also our focus here. 

Usually, 1D KPZ (as well as other growing) interfaces evolve from a flat line of fixed size, $L_0$, driven by a homogeneous deposition flux, so that they are flat, on average. In this case, during the so-called \textit{growth regime} (GR) - when the correlation length $\xi\simeq c t^{1/z}$ is much smaller than the system size ($\xi \ll L_0$) - the asymptotic (1-point) height distribution (HD) of 1D KPZ interfaces is given by the Tracy-Widom \cite{TW} distribution from a Gaussian orthogonal ensemble (GOE), as confirmed in a number of works \cite{Prahofer2000,Calabrese2011,Takeuchi2011,tiago12,healy14,Silvia17}. Furthermore, the (2-point) spatial covariance of the interface is known to be given by the so-called Airy$_1$ process \cite{Sasa2005}.

For curved 1D KPZ interfaces, on the other hand, the HD changes to the TW distribution from a Gaussian unitary ensemble (GUE) \cite{Prahofer2000}, whereas the spatial covariance is related to the Airy$_{2}$ process \cite{Prahofer2002}. GUE fluctuations have been indeed found at the ``central point'' of a number of curved 1D KPZ interfaces with open boundary conditions (BCs), both analytically \cite{Johansson,Prahofer2000,Sasamoto2010,Amir} and numerically \cite{Alves2013,alves13,healy14}. Moreover, GUE HD and Airy$_2$ covariance have been found in some circular KPZ interfaces (evolving radially) on the plane as is the case in experimental interfaces of turbulent phases in liquid crystal films \cite{Takeuchi2010}; and in simulations of the classical Eden model \cite{Eden} starting from a single seed \cite{Alves11}, and of discrete models growing on expanding substrates \cite{Ismael14}; among others \cite{Silvia14,Sidiney18}.

\begin{figure}[!b]
\includegraphics[height=2.5cm]{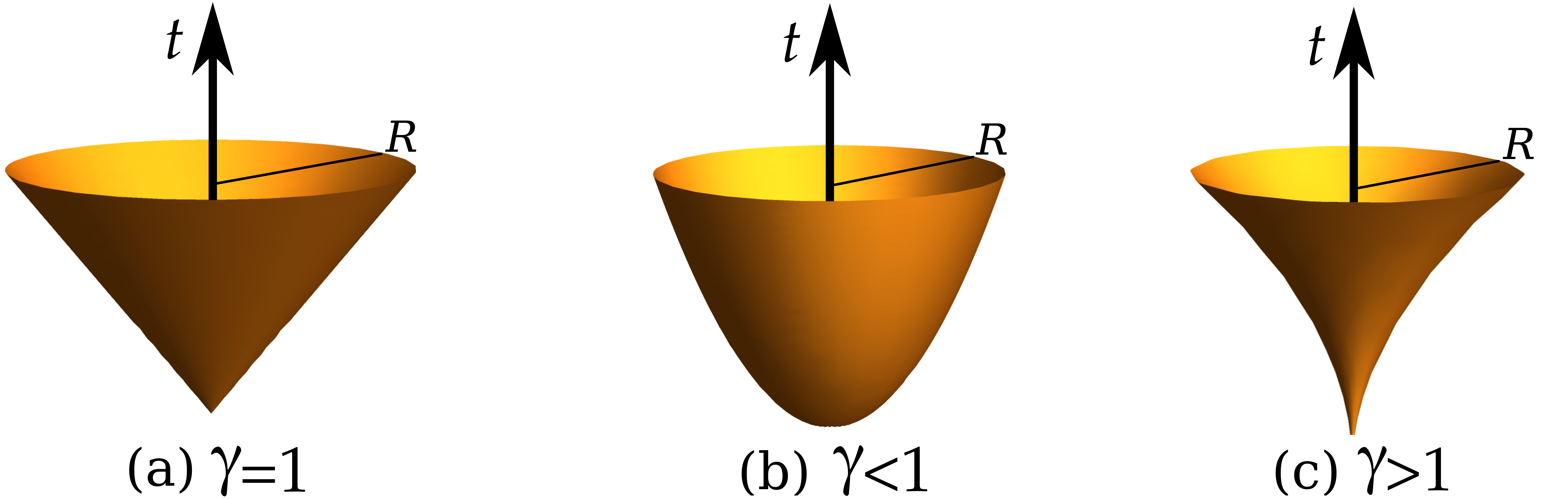}
\caption{Illustration of the temporal evolution of the interfaces in space for different expansion exponents $\gamma$.}
\label{fig1}
\end{figure}

In all these systems, the circular interfaces evolve on the plane, or mimic this situation, so that their (average) perimeters $\left\langle L(t) \right\rangle$ - and so their radii $\left\langle R\right\rangle$ - do increase linearly in time, as also do their mean heights $\left\langle h \right\rangle$ [$\left\langle L(t) \right\rangle \sim \left\langle h \right\rangle \sim t$]. This condition is also satisfied for 1D KPZ interfaces evolving on the surface of conical manifolds, studied recently by Santalla \textit{et al.} \cite{Silvia17}, where again GUE fluctuations were found. On the other hand, if a circular interface evolves on the surface of a non-planar (and non-conical) background space, its size $\left\langle L(t) \right\rangle$ might increase nonlinearly in time while $\left\langle h \right\rangle \sim t$. The KPZ statistics in this very interesting situation, which is so relevant from a practical perspective as the planar case, has never been tackled, for the best of our knowledge. In this work, we investigate this for the case where the interface size enlarges as $\langle L(t) \rangle=L_0+\omega t^{\gamma}$, whereas $\langle h \rangle \sim t$ (see the characteristic surfaces generated in Fig. \ref{fig1}). From extensive kinetic Monte Carlo simulations of several discrete KPZ models a very rich behavior is found in the HDs and covariances as the parameters $L_0$, $\gamma$ and $\omega$ are changed, which substantially generalizes our understanding of 1D KPZ systems.

To demonstrate the universality of our results, we investigate three discrete models which are well-known to belong to KPZ class: the Etching model by Mello {\it et. al} \cite{Mello}, the restricted solid-on-solid (RSOS) model by Kim and Kosterlitz \cite{kk} and the single step (SS) model \cite{barabasi}. In all cases, particles arrive at the deposit at randomly chosen sites and aggregate at a given site $i$ following the rules: \textit{Etching:} $h_i \rightarrow h_i +1$ and, then, $h_{i\pm1} \rightarrow max[h_{i\pm1},h_i -1]$; \textit{RSOS:} $h_i \rightarrow h_i +1$ if $|h_i-h_{i\pm 1}| \leqslant 1$ after deposition; \textit{SS:} $h_i \rightarrow h_i +2$ if $(h_i-h_{i\pm 1}) = 1$ after deposition. The growth starts on a substrate with $L_0$ sites, with $h_i=0$ $\forall$ $i \in [1,L_0]$ for the Etching and RSOS models, and $h_i=1$ (0) for $i$ odd (even) for the SS model. Periodic boundary conditions are used. Following the method from Ref. \cite{Ismael14}, the enlargement of the substrate is implemented by duplicating columns at rate $\delta=dL/dt=\gamma \omega t^{\gamma-1}$, yielding an average substrate size $\left\langle L(t) \right\rangle = L_0+\omega t^{\gamma} $. Depositions and duplications are randomly mixed in a way that in one time unity the average number of duplications (depositions) is equal to $\delta$ ($L$). To do so, at each time step, $\Delta t=1/(L+\delta)$, one deposition is performed with probability $P_L=L/(L+\delta)$ or a column is duplicated with $P_{\delta}=\delta/(L+\delta)$. To conserve the steps in the SS model, a pair of neighbor columns are simultaneously duplicated and, then, $\delta/2$ is used in the expressions for the probabilities and $\Delta t$.

\textit{Results for $\gamma < 1/z$ -} Let us start noticing that for any $\gamma < 1/z = 2/3$ the correlation length increases faster than $L(t)$ and, thus, at a crossover time $t_c$ one must have $\xi \sim L$ and thenceforth the interfaces become completely correlated. In flat systems ($\omega=0$), this gives rise to a \textit{stationary regime} (SR) where the squared interface width [the variance of the HDs - $w_2 = \langle h^2 \rangle_c$] scales with the system size $L_0$ as $w_2 \sim L_0^{2\alpha}$. The consequence of $\xi \sim L$ has also been unveiled recently for ingrowing circular KPZ interfaces \cite{Fukai17,Ismael18}, which corresponds to the case $\gamma=1$ and $\omega <0$ in our framework. For interfaces expanding as $L \sim t^{\gamma}$ we shall have $w_2 \sim t^{2\alpha\gamma}$, as already observed for the KPZ \cite{Pastor2007} and other universality classes \cite{Escudero2}, and confirmed here in Fig. \ref{fig2}a. Note that $\alpha=1/2$ for the 1D KPZ class \cite{KPZ}, leading to $w_2 \sim L \sim t^{\gamma}$.

The KPZ nonlinearity is known to become irrelevant in the SR of flat 1D KPZ interfaces, leading to Gaussian HDs \cite{barabasi}. Figure \ref{fig2}b shows that this is also the case in the non-stationary, but correlated, regime found in our systems, since the cumulant ratios $R=\sqrt{\langle h^2 \rangle_c}/\langle h \rangle$ (the variation coefficient), $S=\langle h^3\rangle_c/\langle h^2\rangle_c^{2/3}$ (the skewness) and $K=\langle h^4\rangle_c/\langle h^2\rangle_c^2$ (the kurtosis) vanish for long times. Here $\langle h^n \rangle_c$ denotes the $n^{th}$ HD's cumulant. As shows Fig. \ref{fig2}d, the spatial covariance $C_S(r,t) \equiv \left\langle \tilde{h}(x,t) \tilde{h}(x+r,t) \right\rangle$, where $\tilde{h} \equiv h - \left\langle h\right\rangle $, is also the same as the one for the SR. This last one was obtained simulating the models with $\omega=0$ in the SR. [See \cite{SupMat} for a discussion about their rescaling.] These results demonstrate that despite their non-stationarity (since $w_2 \sim t^{\gamma}$) 1D KPZ interfaces with size expanding as $L \sim t^{\gamma}$ with $\gamma<1/z$ have asymptotically the same statistics of the stationary regime found in flat (fixed-size) 1D KPZ systems.

At short times there exists a \textit{growth regime} (GR), where $w_2 \sim t^{2\beta}$, whose duration is determined by the initial size $L_0$, as shows Fig. \ref{fig2}a. In fact, we have that $w_2/L(t)^{2\alpha} \times t/L(t)^z$ collapses quite well for a given model, so that the crossover from the GR to SR shall occur when $t_c^{\beta} \sim L(t_c)^{\alpha}$, so that the crossover time satisfies $t_c \sim L_0^z/\left[1 - \omega t_c^{-(1/z-\gamma)} \right]^z$. As shows Fig. \ref{fig2}c, the large $L_0$ (and so $t_c$) is, the closer the HDs become of the GOE distribution before they cross over to the asymptotic correlated regime. Moreover, within the time window where the HDs are close to GOE, the rescaled spatial covariances approach the Airy$_1$ curve, but they move towards the SR covariance for long times  (see Fig. \ref{fig2}d).

\begin{figure}[!t]
\includegraphics[width=4.25cm]{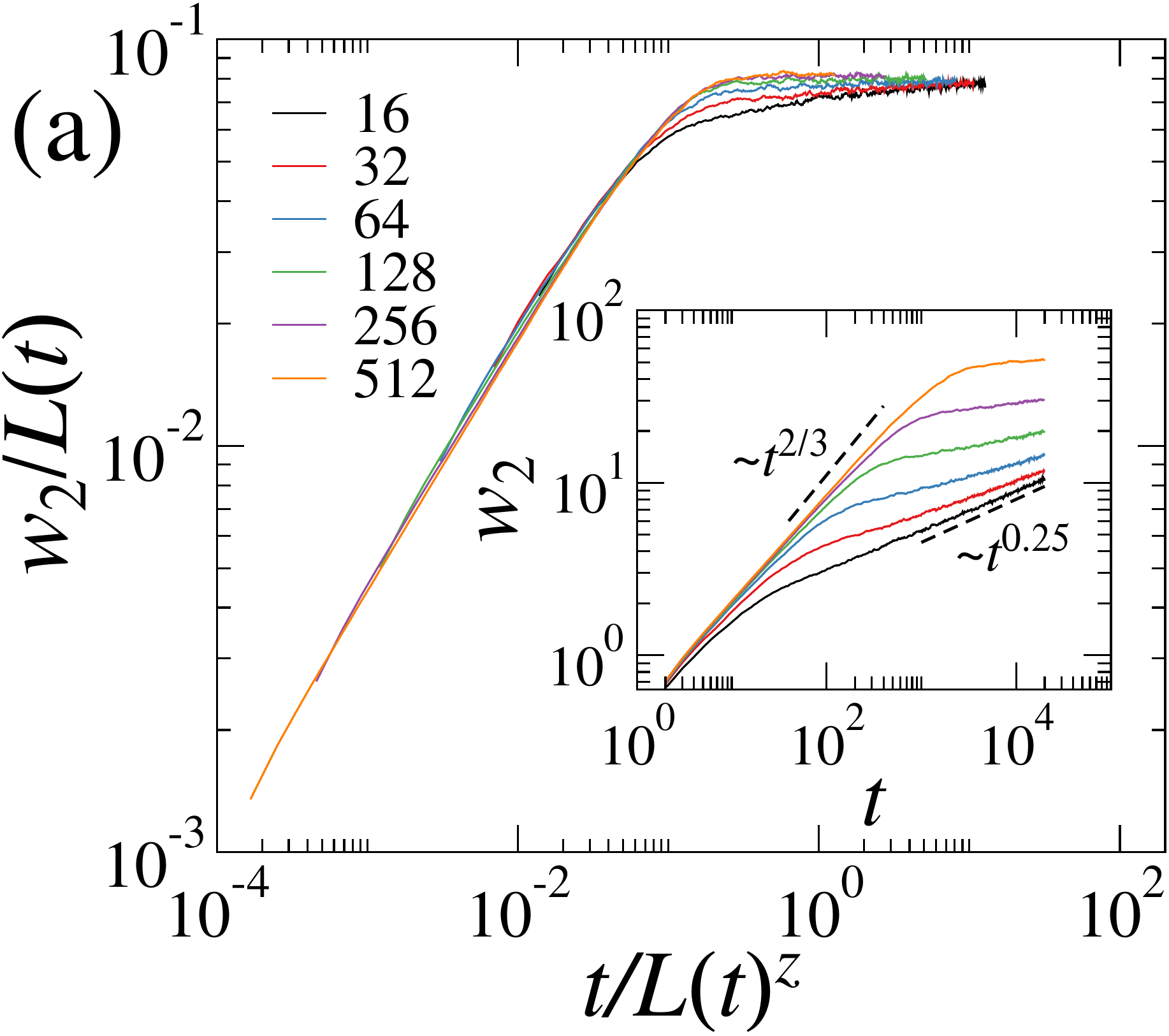}
\includegraphics[width=4.25cm]{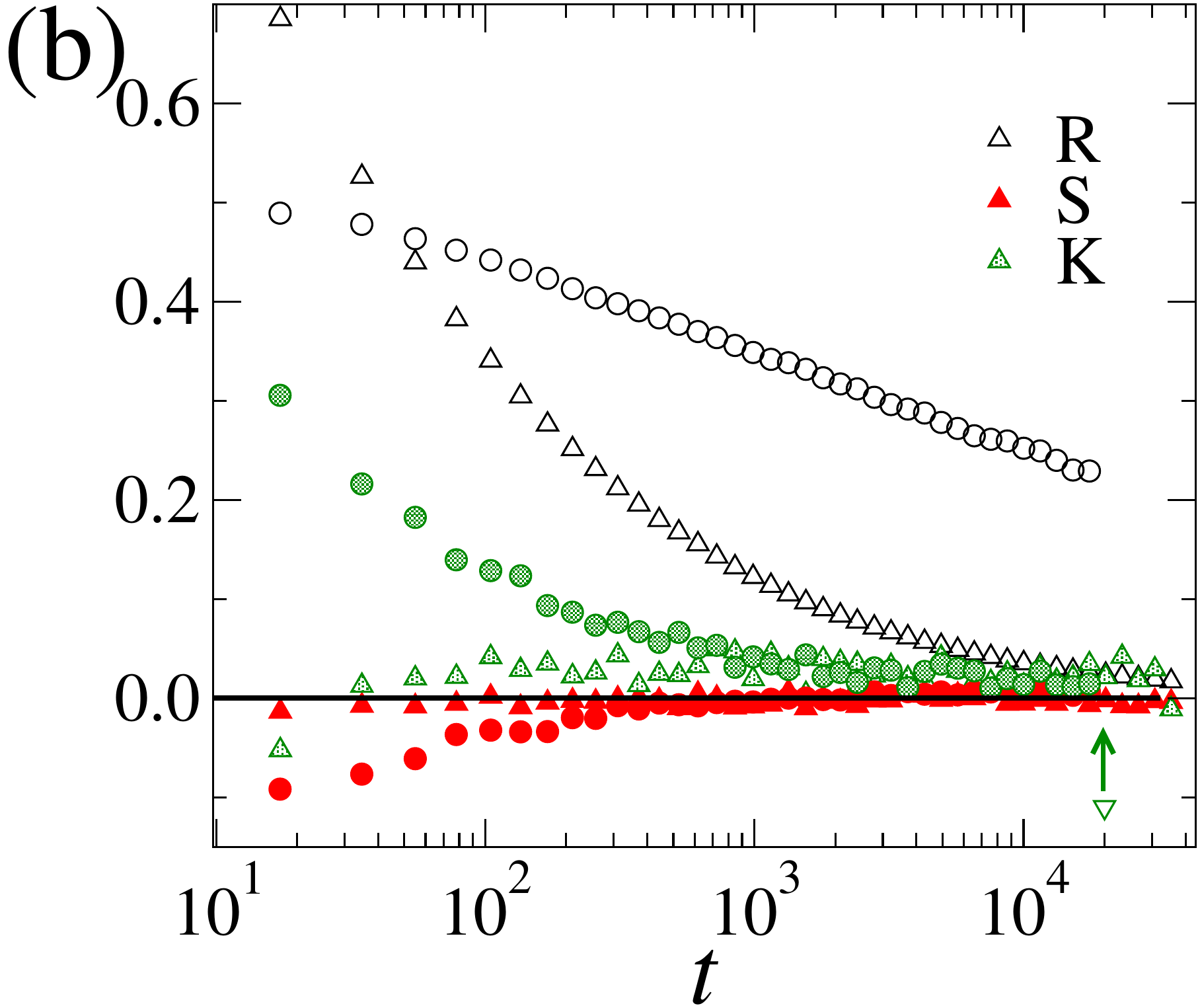}
\includegraphics[width=4.25cm]{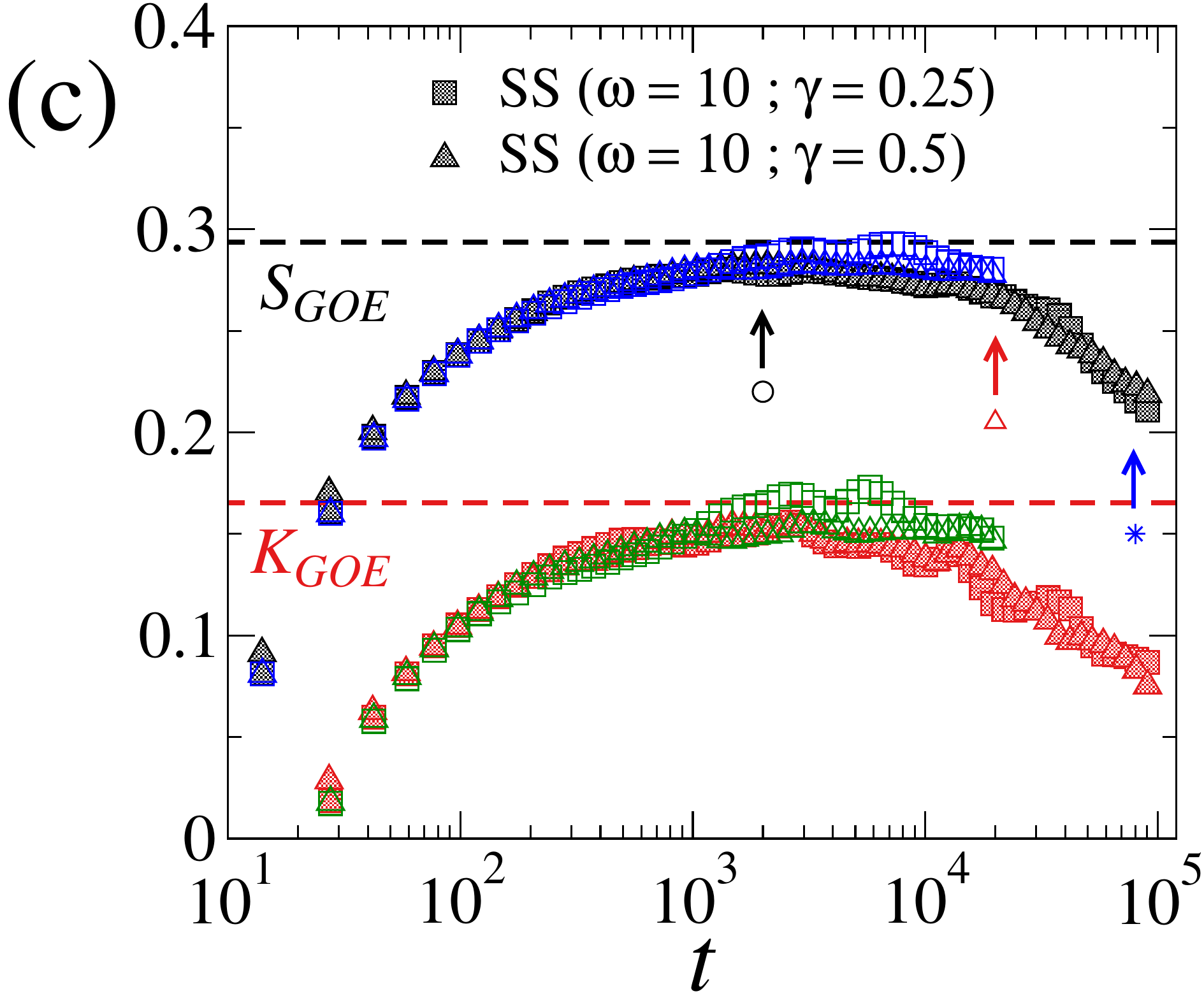}
\includegraphics[width=4.25cm]{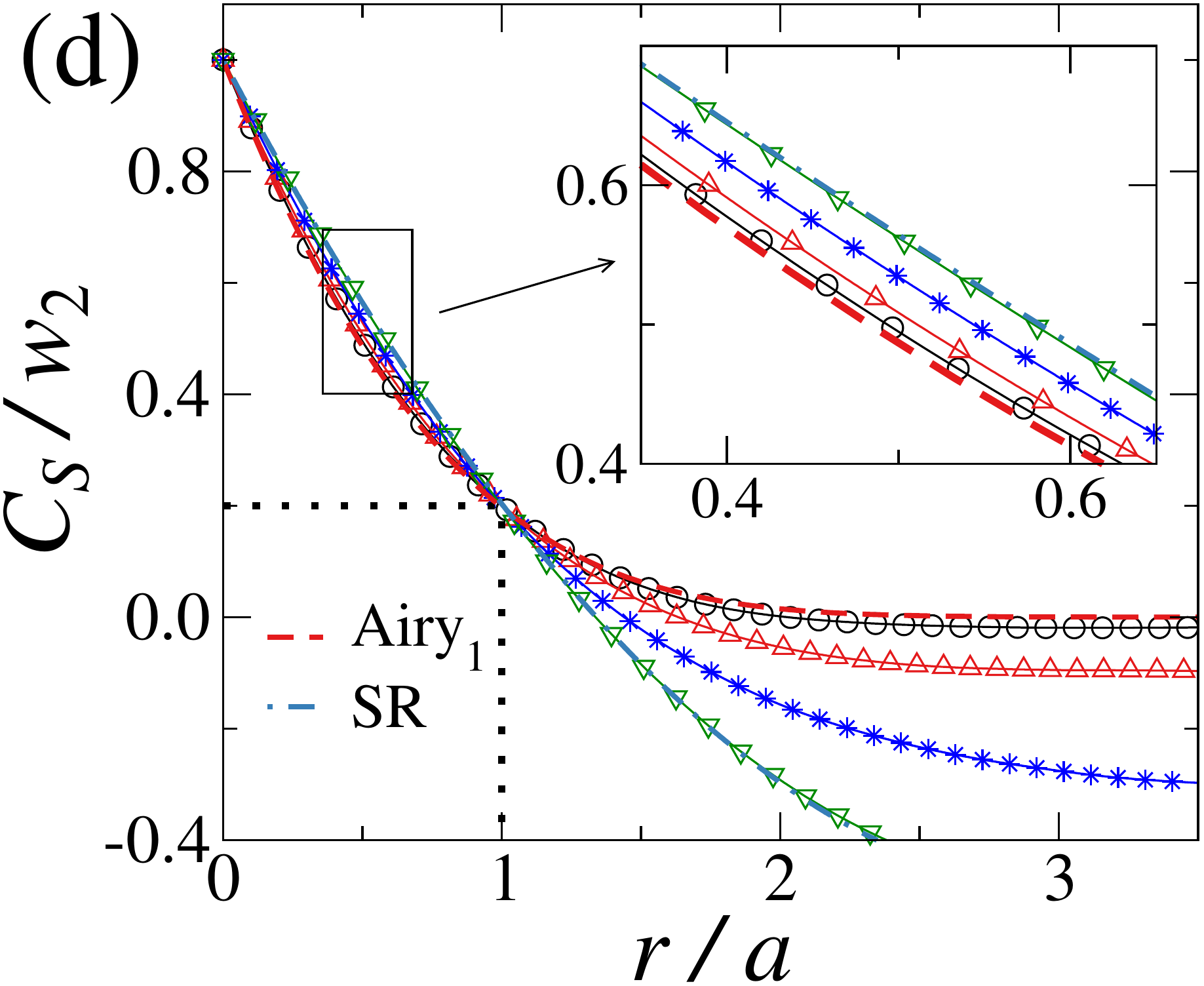}
\caption{(Color online) \textit{Results for $\gamma<1/z$.} (a) Collapse of $w_2 \times t$ curves for the SS model, with $\omega=10$, $\gamma=1/4$ and the several values of $L_0$ indicate. The insertion shows the same data non-rescaled, where the dashed lines have the indicated slopes. Temporal evolution of the cumulant ratios (b) $R$, $S$ and $K$ for the SS (with $\gamma = 0.25$ - triangle) and etching (with $\gamma = 0.5$ - circle) models, with $L_0=4$ and $\omega=20$; and (c) $S$ and $K$ for the SS model with $L_0=20000$ (full) and $L_0=40000$ (open symbols). (d) Rescaled spatial covariances for the SS model, for the same parameters and times indicated by the arrows in Figs. \ref{fig2}b and \ref{fig2}c. The colors in (and symbols with) such arrows are the same used in (d).}
\label{fig2}
\end{figure}

Therefore, overall the statistics of interfaces expanding slower than $\xi$ is the same as in the flat case. As demonstrated in Refs. \cite{Ismael14,Ismael18}, the flat statistics appears even in systems expanding faster than $\xi$, at short times, provided that $L_0$ is large enough. Hence, since our focus here is in the asymptotic behavior, to avoid undesirable crossover effects yielded by $L_0$ in all results that follows we will set $L_0=4$.

\textit{Results for $\gamma = 1/z$ -} When both $L(t)$ and $\xi(t)$ increase following the same scaling, we have a special situation where the parameter $\omega$ is expected to play a key role. Noteworthy, it does not matter whether such system is correlated or not, $w_2$ shall increase as in the growth regime, namely, $w_2 \sim t^{2 \beta}$, with $\beta=1/3$, since in the correlated case $w_2 \sim t^{2\alpha \gamma} \sim t^{2\alpha/z}\sim t^{2\beta}$. This is indeed confirmed in Fig. \ref{fig4}a (see also Refs. \cite{Pastor2007,Escudero2}).

Figures \ref{fig3}a and \ref{fig3}b respectively show the temporal evolution of the skewness $S$ and kurtosis $K$ of the HDs for the Single Step (SS) model and several values of $\omega$. After a transient, $\omega$-dependent plateaus are observed in these quantities, indicating that the asymptotic regime has been attained. Similar results are found for all models investigated here. For very large $\omega$'s, the curves of $S$ and $K$ surpass the GUE values, but they do not have clear plateaus and they seem to converge towards the GUE values from above. This behavior is certainly related to the GOE-GUE crossover mentioned above, since a large $\omega$ makes $L(t)$ large at very short times, when the interfaces are still very smooth, having an effect similar to a large $L_0$. Once the GOE-GUE crossover is quite slow (see Ref. \cite{Ismael14}), it is pretty hard to attain the GUE regime for $\omega \gg 1$.

The asymptotic values of $S$ and $K$ are displayed in Fig. \ref{fig3}c as function of the ratio $\mathcal{R}\equiv \omega/c \simeq L(t)/\xi(t)$. The amplitudes of $\xi$ were obtained from data in Ref. \cite{Ismael14}, being $c_{SS} = 1$, $c_{RSOS} = 0.783$ and $c_{etch} = 1.663$. The data collapse in Fig. \ref{fig3}c demonstrates that for $\gamma=1/z$ the KPZ HDs are determined solely by the ratio $\mathcal{R}$. Significantly, there exist a class of distributions continuously interpolating between Gaussian (for $\mathcal{R} \rightarrow 0$) and GUE (for $\mathcal{R} \gg 1$). As an aside, we notice that a class of continuously changing HDs was recently found in 1D KPZ systems with correlated noise \cite{Kardar16}. In this case, the HDs interpolate between GOE and Gaussian as the strength of the noise correlation increases, but this change is accompanied by a change in the scaling exponents and, thus, it could be expected. Another class of asymptotic HDs continuously interpolating between GOE and Baik-Rains \cite{Baik} statistics has also been found for 1D KPZ systems with spatially homogeneous random initial conditions \cite{Ferrari18}, as their diffusion coefficient changes. In contrast with our system, in both of the cases above the interfaces are macroscopically flat and thus have GOE fluctuations as one limit, while here we find GUE. 

Once one has $w_2 \sim t^{2\beta}$, regardless the parameter $\mathcal{R}$, the height at a given point of the interfaces shall follow the KPZ ansatz \cite{Krug1992,Prahofer2000}
\begin{equation}
h=v_{\infty}t+s_{\lambda}(\Gamma t)^{\beta} \chi + \dots,
\label{ansatz}
\end{equation}
where $v_{\infty}$, $s_{\lambda}(=\pm1)$ and $\Gamma$ are system-dependent parameters, $\beta=1/3$, and $\chi$ is a random variable fluctuating according to GUE (GOE) in curved (flat) 1D KPZ interfaces. For $\gamma=1/z$, one should have $\chi = f(\mathcal{R},\chi_{GUE},\chi_{Gauss})$, with $f(\mathcal{R} = 0) = \chi_{Gauss}$ and $f(\mathcal{R} \gg 1)=\chi_{GUE}$. 

In the way between Gaussian (where $S=K=0$) and GUE (where $S=0.2241$ and $K=0.0934$) the increasing cumulant ratios in Fig. \ref{fig3}c have to pass through the values for the TW distribution from a Gaussian symplectic ensemble (GSE), which are $S=0.1655$ and $K=0.0492$ \cite{tw2}. Interestingly, our results strongly indicate that this occurs at the same point $\mathcal{R} \approx 10$ for both $S$ and $K$, so that for some value of $\mathcal{R}$ near 10 the HDs are given by the GSE TW distribution. We remark that, so far, GSE statistics have been found only at one point of 1D KPZ interfaces with very special conditions, namely, at the boundary of half-space KPZ interfaces, when they have sources or constraints \cite{Prahofer2000,Sasa07,Sasa04,LeDoussal12,Borodin16}. Numerical evidence of GSE HDs have also been found recently at the origin of droplet KPZ interfaces expanding with different (but constant) speeds in the left and right halves of the space \cite{Ito18}. Therefore, the evidence of GSE found here through a fine-tuning balance between $L$ and $\xi$ suggests that the role of this distribution within the 1D KPZ class is much more general and it can set the fluctuations of the whole interface, rather than a single special point.

\begin{figure}[!t]
\includegraphics[width=4.25cm]{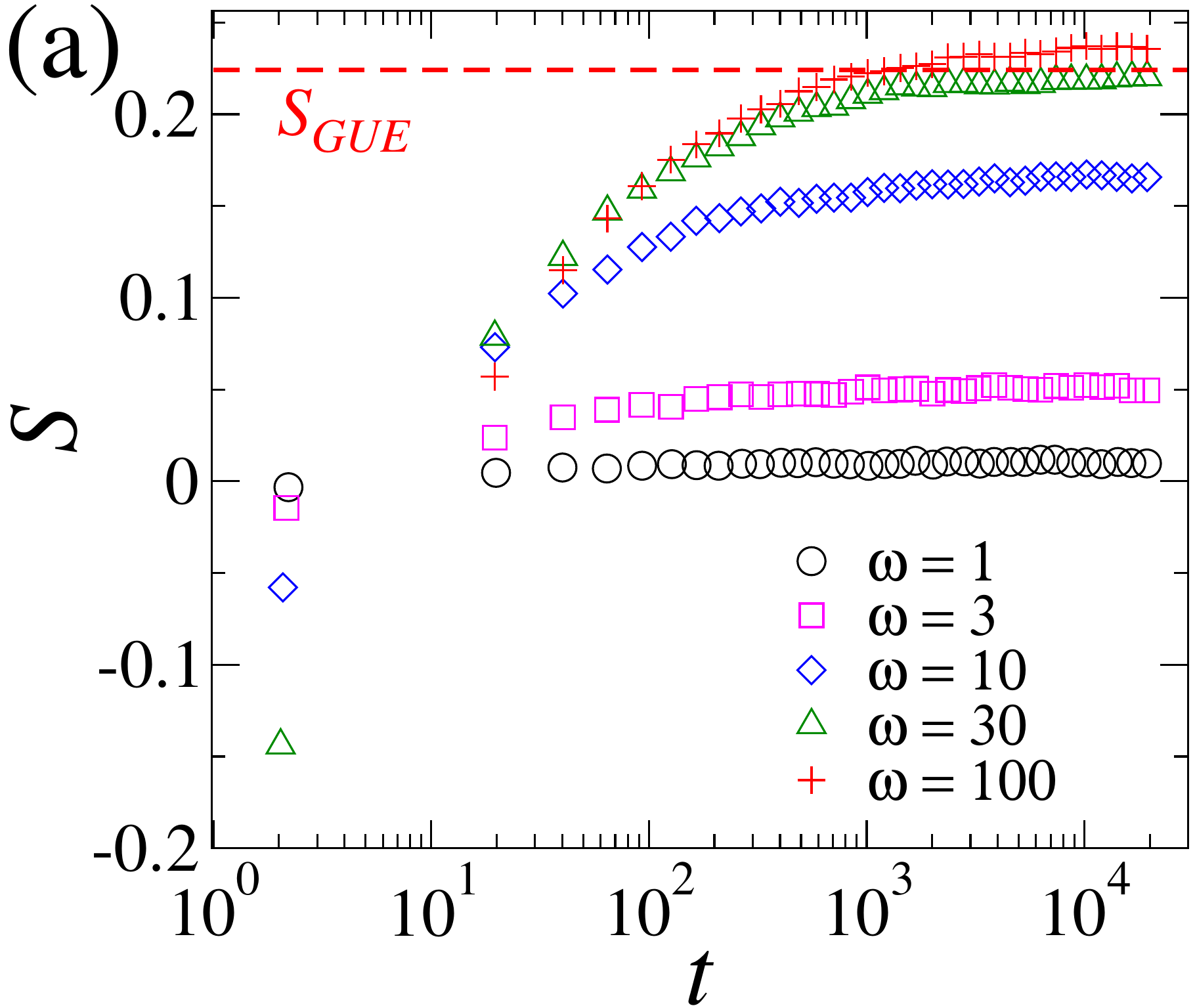}
\includegraphics[width=4.25cm]{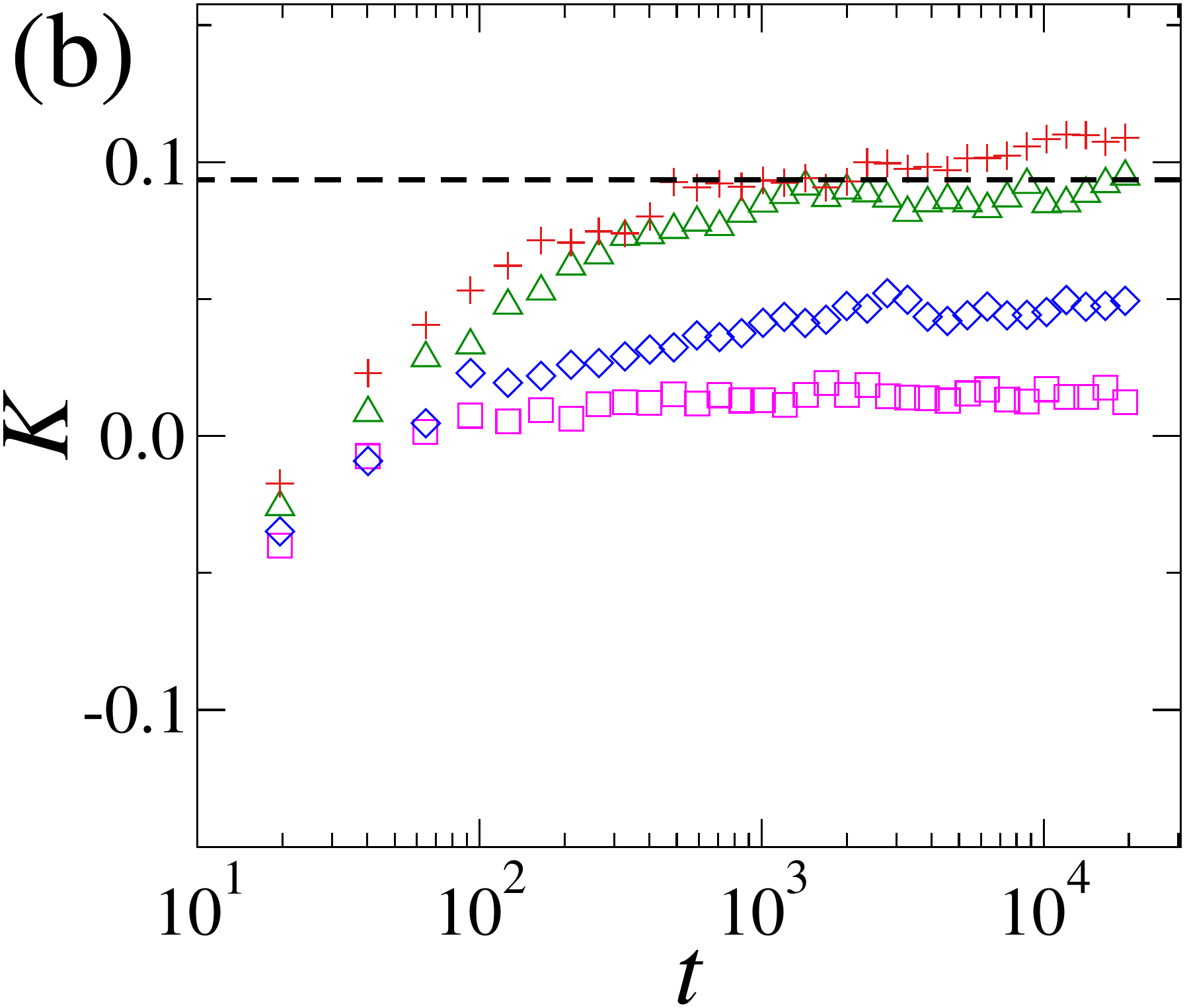}
\includegraphics[width=4.25cm]{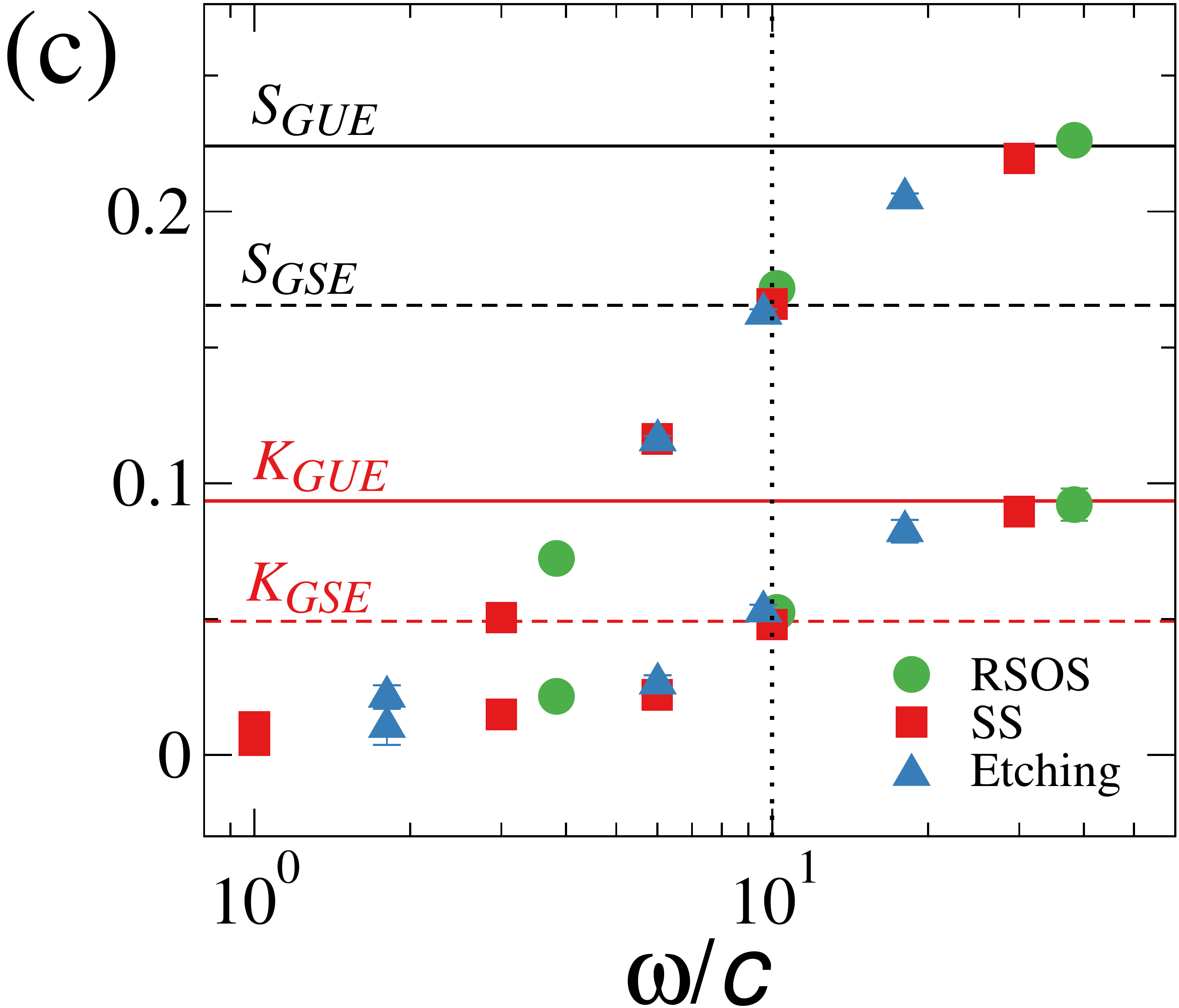}
\includegraphics[width=4.25cm]{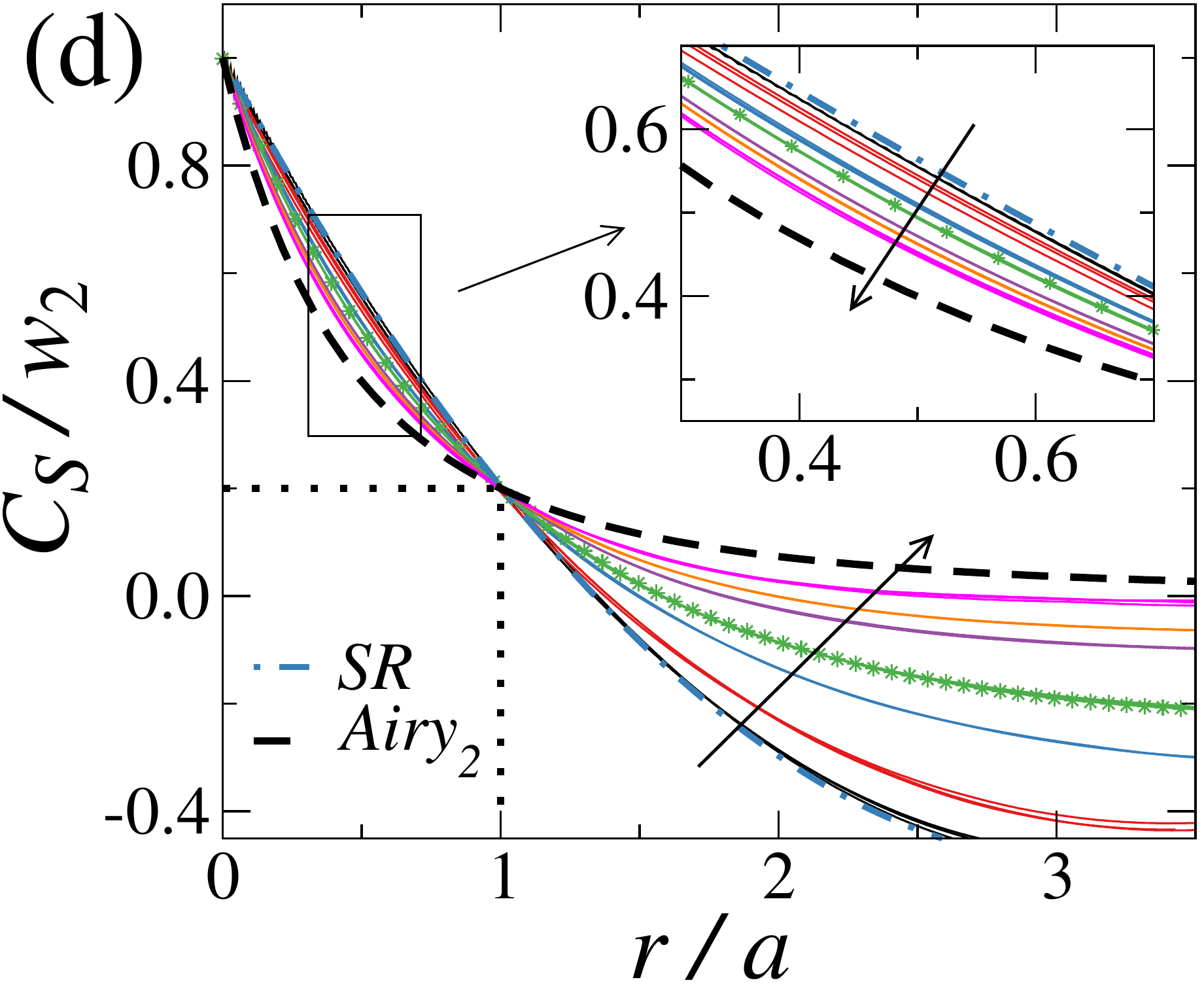}
\caption{(Color online) \textit{Results for $\gamma = 1/z$.} Temporal evolution of (a) the skewness $S$ and (b) the kurtosis $K$ for the SS model and the values of $\omega$ indicated. (c) Asymptotic values of $S$ and $K$ as function of the ratio $\mathcal{R} =\omega/c$ for the three models investigated. (d) Rescaled spatial covariance for the SS model and several values of $\omega$ in the range $[1,100]$, whose increasing is indicate by the arrows. For each $\omega$, curves for three different times are shown, which approximately collapse onto single curves. The curve with symbols is for $\mathcal{R} = 10$.}
\label{fig3}
\end{figure}

Figure \ref{fig3}d presents the rescaled spatial covariance for several values of $\omega$ (and so of $\mathcal{R}$) for the SS model. A similar behavior is found for the other models. As one can see, the continuous variation in the HDs is accompanied by continuously changing covariances and they interpolate between the SR one (for $\mathcal{R} \rightarrow 0$) and the Airy$_2$ (for $\mathcal{R} \gg 1$). The covariance for $\mathcal{R} =10$ is located in between. It is highlighted in Fig. \ref{fig3}d and may be seem as the partner of the GSE distribution. Thereby, when $\gamma=1/z$, both 1-point and 2-point statistics smoothly change from the one for completely correlated systems, to the one for the circular interfaces evolving on the plane.

\textit{Results for $\gamma > 1/z$ -} Finally, we consider the case where $L$ increases faster than $\xi$, so that the system is always in the GR and thus $w_2 \simeq B t^{2/3}$. This is indeed confirmed in Fig. \ref{fig4}a, where we see that while the exponent $\beta$ is independent of $\gamma$ and $\omega$, the scaling amplitude $B$ varies with these parameters. From plots of $w_2/t^{2 \beta} \times t$ (see the insertion in Fig. \ref{fig4}a), we estimate $B$ in the limit $t\rightarrow\infty$, whose variation with $\gamma$ and $\omega$ is presented in \cite{SupMat}. In general, $B$ increases with both $\gamma$ (for fixed $\omega$) and $\omega$ (with $\gamma$ fixed), so that the fast the substrates expand, the large the variance of the HDs for a given time (and model) is. This seems to be related to the fact that several duplications occur at sites with heights very different from $\langle h \rangle$ and, so, they lead to an increase in $B$. Notwithstanding, this seems to be a general feature of expanding interfaces, rather than an effect of our method.

Figure \ref{fig4}b shows the temporal variation of the cumulant ratios $R$, $S$ and $K$ for the three studied models and several parameters ($\gamma$ and $\omega$). In all cases, the ratios converge to the GUE values, showing that the 1D KPZ HDs are always given by GUE for any $\gamma > 1/z$. Therefore, the GUE statistics is not a peculiarity of 1D KPZ systems evolving on the plane (or on a conical surface), but a rather general asymptotic behavior of expanding interfaces which do not become completely correlated when $t \rightarrow \infty$.

In face of the universal GUE HDs, the variation in $B$ and the KPZ ansatz (Eq. \ref{ansatz}), we are lead to conclude that $\Gamma=\left( B/\langle \chi^2 \rangle_c \right)^{1/2\beta}$ is a function of $\gamma$ and $\omega$ (see \cite{SupMat}). Furthermore, the KPZ ansatz as presented in Eq. \ref{ansatz} is not complete, since additional corrections are expected on it. Beyond the well-known additional constant correction $\eta$ \cite{Takeuchi2010,Sasamoto2010,Alves11,Ismael14}, in our system the duplication of columns yields a correction $\zeta t^{1-\gamma}$, whose derivation is presented in \cite{SupMat}, where a numerical confirmation of it is also shown. Thereby, one must have 
\begin{equation}
h=v_{\infty}t+s_{\lambda}(\Gamma t)^{\beta} \chi + \eta + \zeta t^{1-\gamma} +\dots,
\end{equation}
where $\eta$ and $\zeta$ are (in principle) stochastic variables. Note that for $\gamma > 1$, the correction $\zeta t^{1-\gamma}$ becomes negligible at long times, while for $1/z < \gamma \leqslant 1$ it might be relevant. This explains the slow (fast) $R$-convergence observed in Fig. \ref{fig4}b for small (large) $\gamma$'s. We remark that the asymptotic growth velocities $v_{\infty}$  - whose values can be found in \cite{Ismael14} - are not affected by the duplications. 

\begin{figure}[!t]
\includegraphics[width=4.25cm]{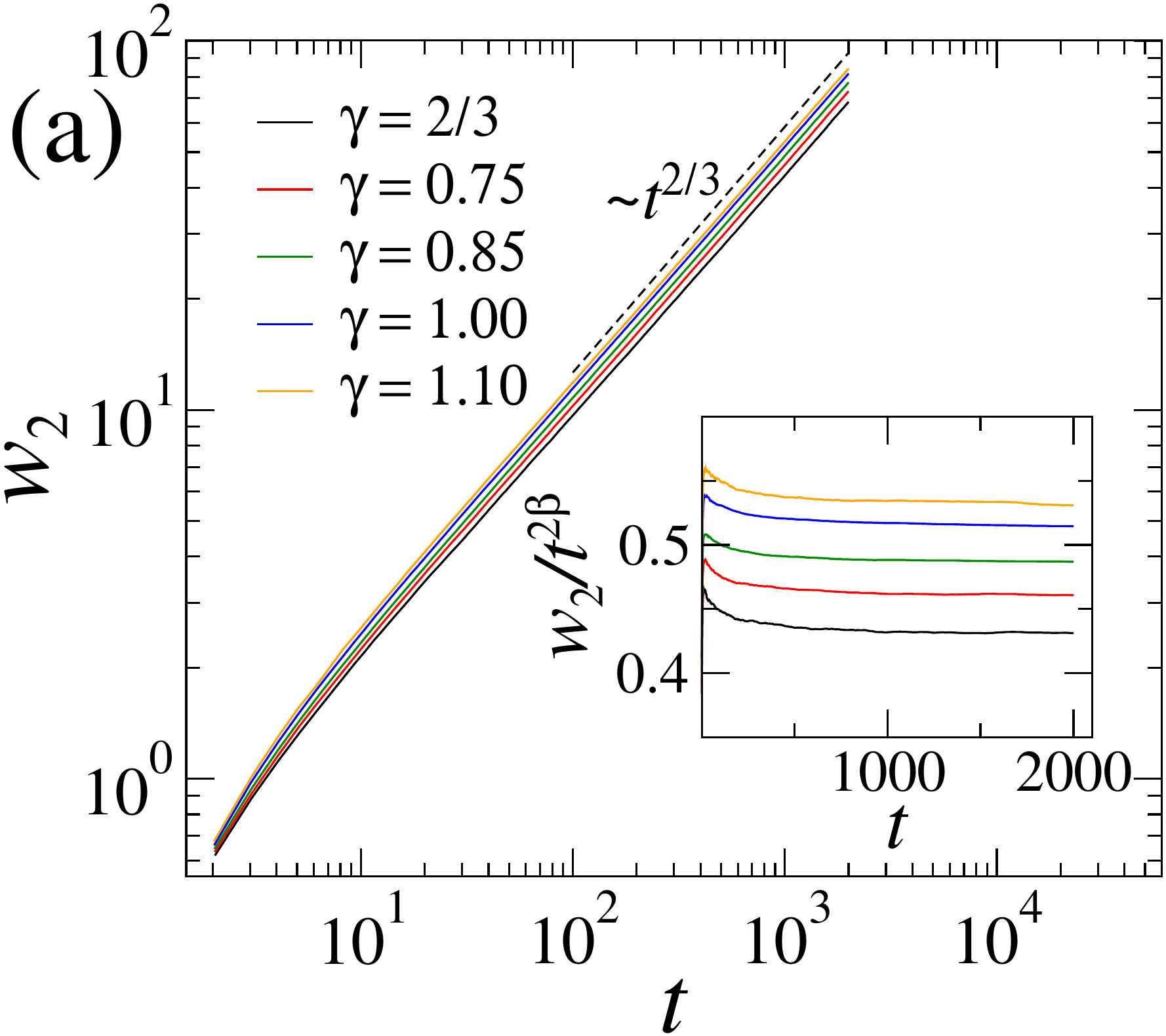}
\includegraphics[width=4.25cm]{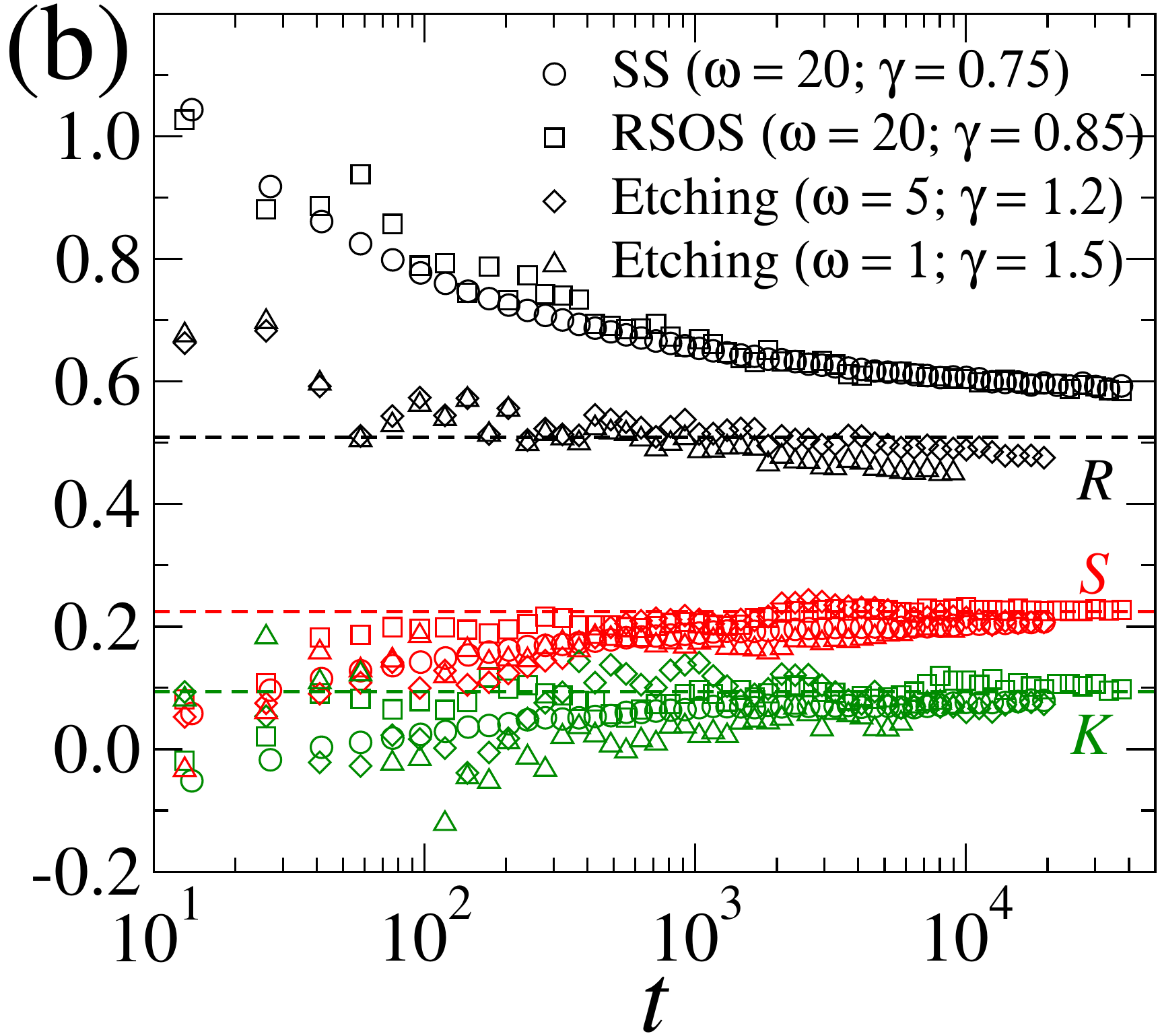}
\includegraphics[width=4.25cm]{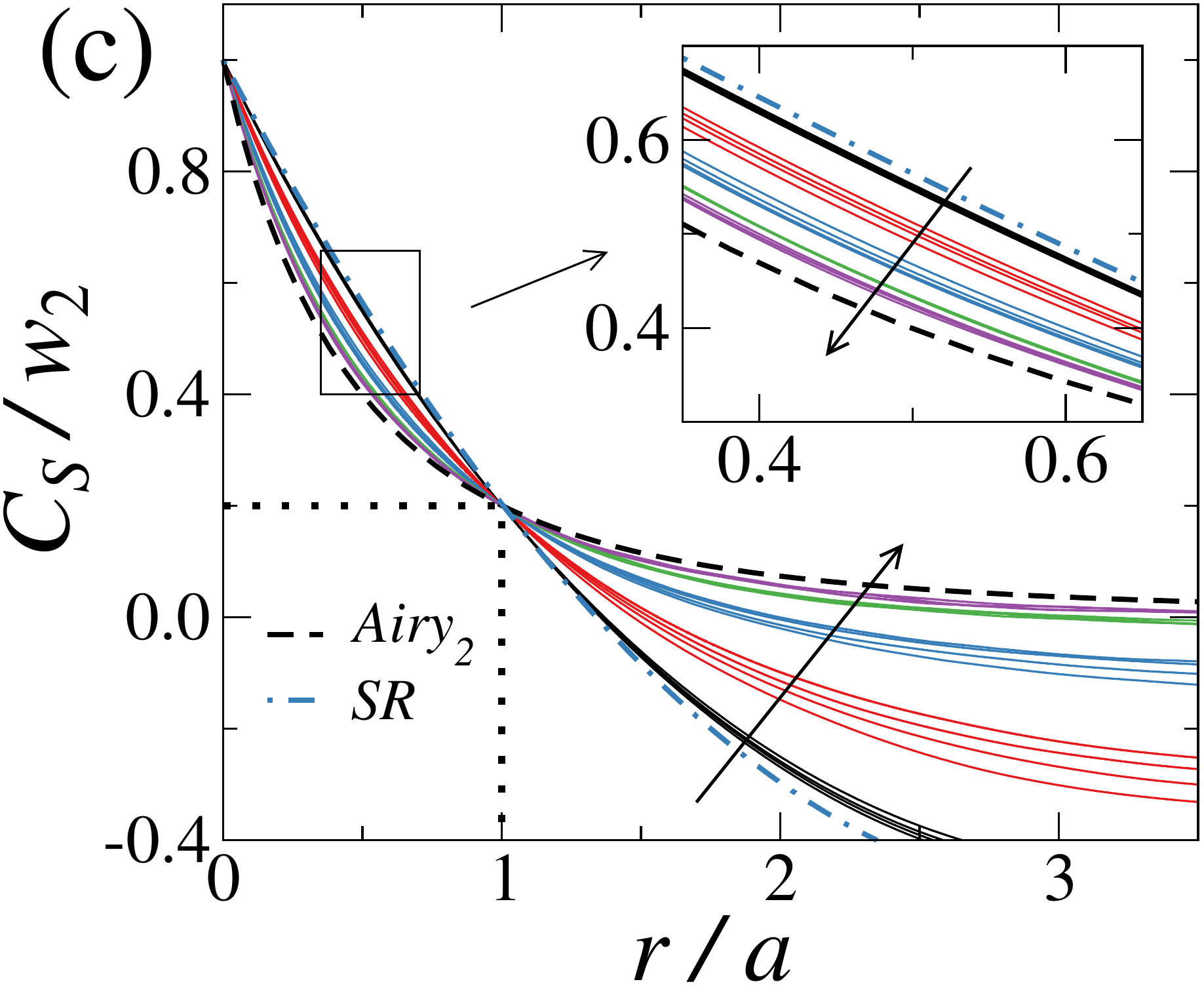}
\includegraphics[width=4.25cm]{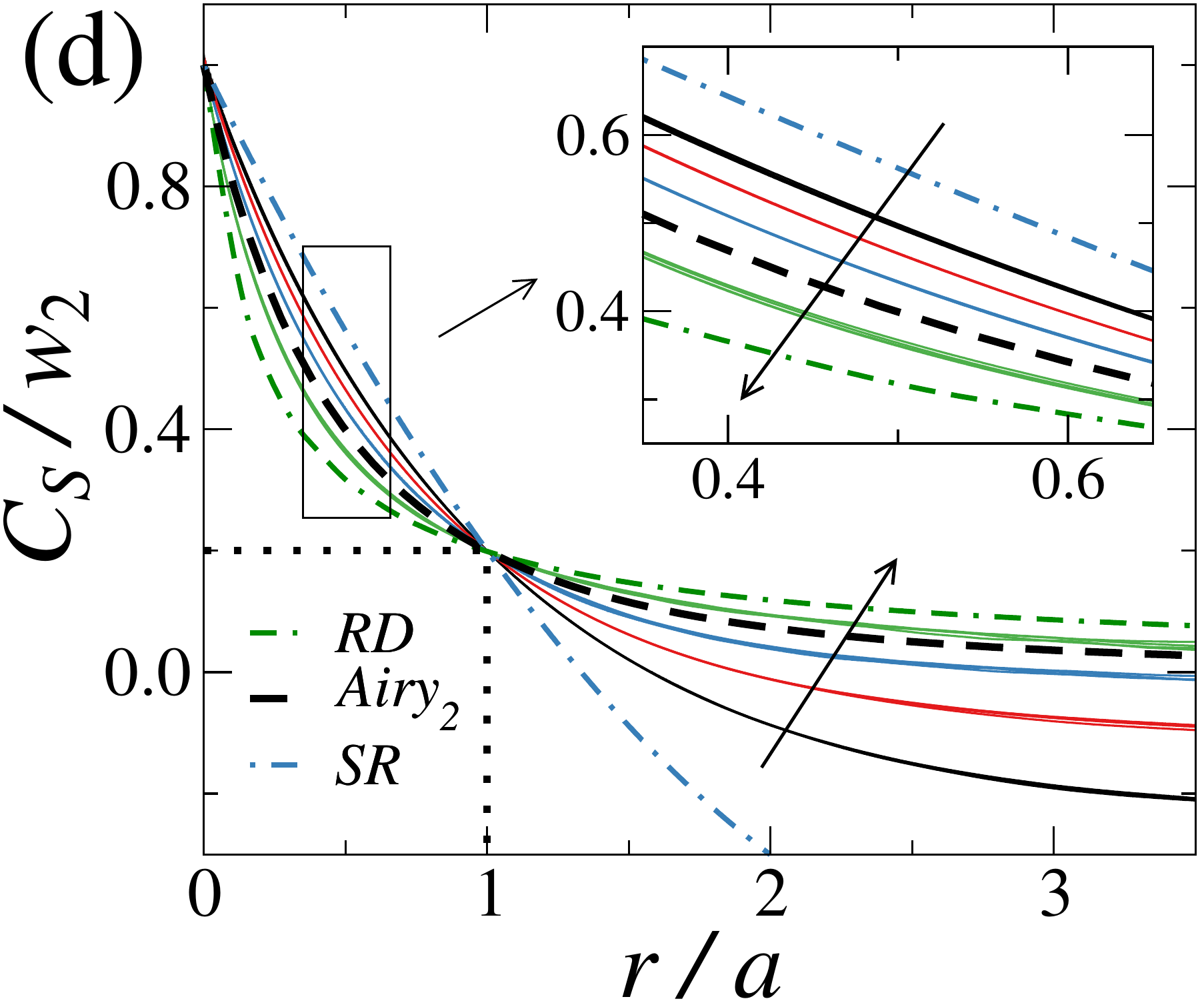}
\caption{(Color online) \textit{Results for $\gamma>1/z$.} (a) Squared interface width $w_2$ versus time for the SS model, with $\omega=20$ and several $\gamma$'s. The insertion shows the same data, but with $w_2/t^{2\beta}$ in vertical axis. (b) Temporal evolution of the cumulant ratios $R$, $S$ and $K$ for the models and parameters indicated. The dashed lines represent the GUE values. Rescaled spatial covariances for (c) $\gamma=0.85$ and several values of $\omega \in [0.25,30]$ and (d) $\omega=10$ and several values of $\gamma \in [2/3,1.2]$, for the SS model. For each set of parameters, data for four times $t\in [4000,160000]$ are shown. The arrows indicate the direction of increasing $\omega$ (and time) in (c) and of increasing $\gamma$ in (d).}
\label{fig4}
\end{figure}

The robustness of the GUE HDs leads us immediately to inquire if the same occurs with the spatial covariance and, interestingly, the answer is negative. Figures \ref{fig4}c and \ref{fig4}d show spatial covariances for $\gamma=0.85$ and several $\omega$'s and for $\omega=10$ and several $\gamma$'s, respectively. In both cases, there exist a clear dependence in the rescaled $C_S$ curves with these parameters. In Fig. \ref{fig4}c we see that for small $\omega$'s such curves suffer from severe finite-time corrections, but no evidence exists that they would converge to the same asymptotic curve. Instead, our data strongly suggests that the (rescaled) asymptotic covariance is a continuous function of $\omega$, for a given $\gamma$, presenting two different behaviors for $\gamma<1$ and $\gamma > 1$. In the former case, the covariances interpolate between the stationary curve when $\omega \rightarrow 0$ and the Airy$_2$ curve for $\omega \gg 1$. Similarly, for a fixed $\omega$, we also find a continuous variation in the covariances with $\gamma$, which only coincide with the Airy$_2$ curve when $\gamma =1$ (see Fig. \ref{fig4}d). This suggests that, for this sublinear expansion of the system, the 2-point spatial statistics carry some characteristics of the one for the correlated (SR) case. In fact, since $\langle L(t) \rangle$ is not expanding much faster than $\xi$, this suggests that the system stays in a kind of crossover state between a pure GR and a pure SR. On the other hand, for a given $\gamma>1$, by increasing $\omega$ and/or the time the covariances do not approach the Airy$_2$ curve, but instead they move in the opposite direction. This can be explained by the fact that now $\langle L(t) \rangle$ is increasing faster than $\langle h \rangle$ and, thus, the spatial correlations yielded by the expansion of the system dominates the KPZ ones coming from the deposition process. To confirm this, we have calculated (numerically) the covariance of interfaces from a random deposition (RD) process \cite{barabasi} expanding as $L \sim t^{\gamma}$ (see \cite{SupMat}), which is also shown in Fig. \ref{fig4}d and, indeed, the KPZ curves slightly move towards the RD one as $\omega$ increases. 

\begin{figure}[!t]
\includegraphics[width=9.cm]{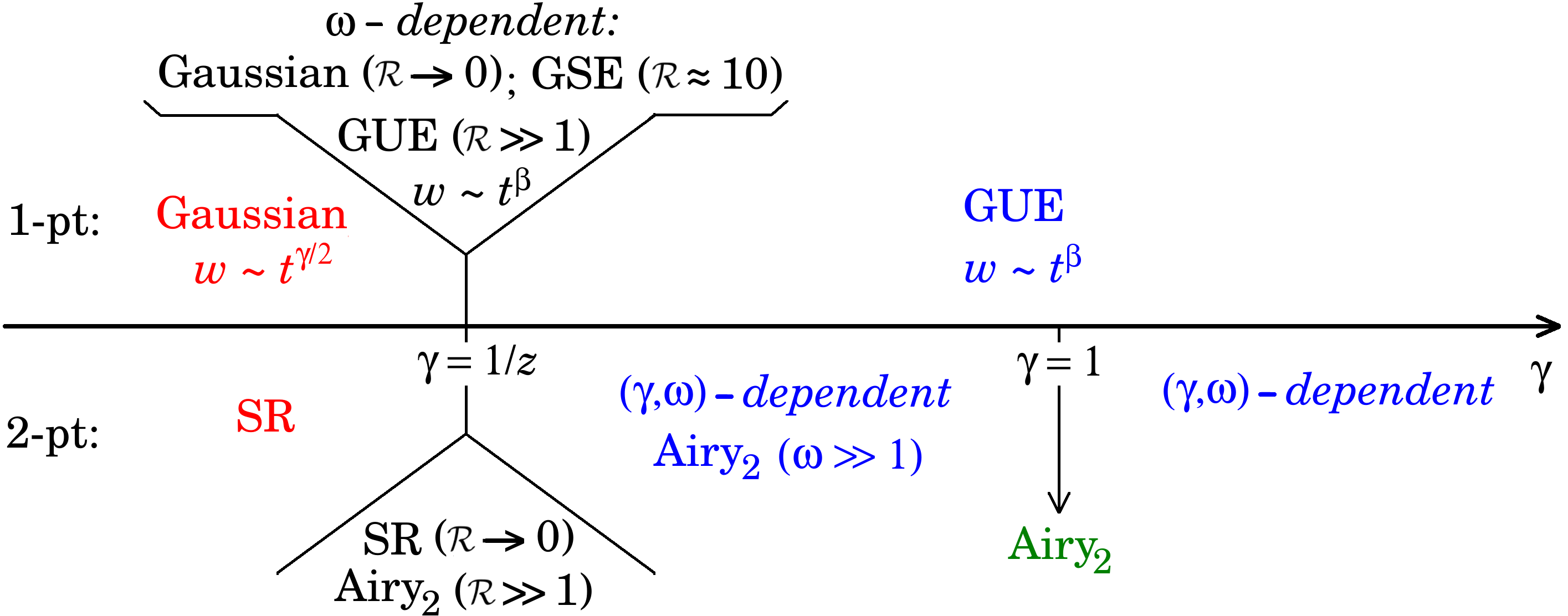}
\caption{(Color online) Sketch of the asymptotic 1-point (above) and 2-point (below the axis) statistics of 1D KPZ interfaces expanding as $L \sim \omega t^{\gamma}$.}
\label{fig5}
\end{figure}

In conclusion, we have demonstrated that the statistical behavior of 1D KPZ interfaces with (average) height and perimeter evolving differently in time - which is the case in interfaces evolving out of the plane on non-conical surfaces - is quite rich, as summarizes Fig. \ref{fig5}. For $\gamma<1/z$, the asymptotic statistics is the same as the one for the stationary regime, where HDs are Gaussian, even though the interface width is still increasing as $w \sim t^{\alpha\gamma}$. For $\gamma=1/z$, a class of asymptotic KPZ distributions exists, which interpolates between Gaussian (for $\mathcal{R} \simeq L/\xi \rightarrow 0$) and GUE (for $\mathcal{R} \gg 1$). For $\mathcal{R} \approx 10$, they seem to agree with GSE TW distribution. We stress that this finding can be appealing from a perspective of investigating the GSE distribution experimentally because in our interfaces all points are statistically equivalent. This is an important advance over other (known) possible setups, as the one proposed in Ref. \cite{Ito18}, where only a single or a few points of the interface are expected to fluctuate according to GSE. For $\gamma>1/z$, the GUE HDs are universal, while the spatial covariances are ($\gamma,\omega$)-dependent and only agree with Airy$_2$ for large $\omega$, for a given $\gamma<1$, or in the usual $\gamma=1$ case. Such results, showing that the 1-point statistics is much more robust than the 2-point spatial one, shall motivate new theoretical works, trying to generalize the existing analytical results (for planar or conical spaces) to the more general case analyzed here. Furthermore, our results might be confirmed numerically within the framework of Ref. \cite{Silvia17}, considering KPZ interfaces evolving on non-conical manifolds. This points also the way to realize this experimentally.

\acknowledgments

This work is supported in part by CNPq, CAPES, FAPEMIG and FAPERJ (Brazilian agencies).

\bibliography{library}

\newpage

\newpage

{\huge Supplemental Material}

\author{I. S. S. Carrasco$^{(a,b)}$}
\email{ismael.carrasco@ufv.br}
\author{T. J. Oliveira$^{(a)}$}
\email{tiago@ufv.br}
\affiliation{$(a)$ Departamento de F\'isica, Universidade Federal de Vi\c cosa, 36570-900, Vi\c cosa, Minas Gerais, Brazil \\
$(b)$ Instituto de F\'isica, Universidade Federal Fluminense, 24210-340, Niter\'oi, Rio de Janeiro, Brazil}

\maketitle

\section{Variation of the interface width scaling amplitude}
\label{secamp}

\begin{figure}[!h]
\includegraphics[width=7.cm]{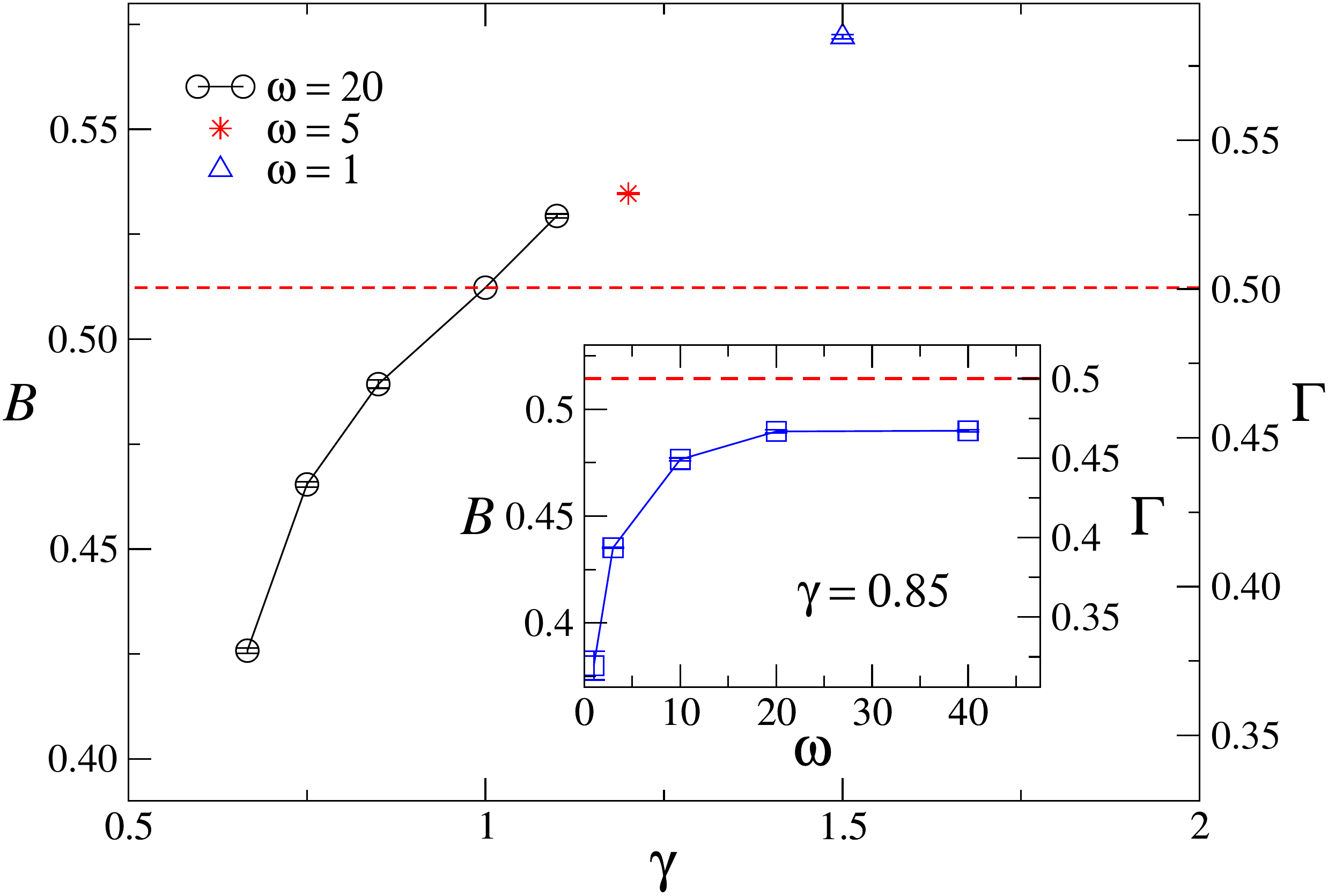}
\caption{Variation of interface width amplitude $B$ (left) and the corresponding $\Gamma$ (right vertical axis) with the exponent $\gamma$ (main plot) and with the parameter $\omega$ (insertion) for the SS model. The dashed horizontal line indicates the exact value of $\Gamma$ (which is $\Gamma=1/2$) for this model in flat (fixed-size) substrates.}
\label{figS1}
\end{figure}

Once the squared interface width scales as $w_2 \simeq B t^{2\beta}$ in 1D KPZ interfaces, the amplitude $B$ can be obtained from the long time limit (when corrections to this scaling relation shall become negligible) of the ratios $w_2/t^{2\beta}$, which are shown in the insertion of Fig. 4a in the main manuscript. The values of $B$ estimated in this way for the SS model are depicted in Fig. S\ref{figS1}. Similar behavior is found for the other models. Thereby, in general, $B$ increases with both $\gamma$ (for fixed $\omega$) and $\omega$ (with $\gamma$ fixed). 

The convergence of the cumulant ratios to the GUE values in Fig. 4b of the main manuscript does not let room for doubt that the HDs are given by GUE, when $\gamma>1/z$. Henceforth, with the KPZ ansatz:
\begin{equation}
h=v_{\infty}t+s_{\lambda}(\Gamma t)^{\beta} \chi + \dots, 
\label{eqansatz}
\end{equation}
bearing in mind, which means that $w_2 \simeq (\Gamma t)^{2\beta} \langle \chi^2 \rangle_c$, we may conclude that $\Gamma=\left( B/\langle \chi^2 \rangle_c \right)^{1/2\beta}$ is also an increasing function of $\gamma$ and $\omega$, assuming that $\langle \chi^2 \rangle_c = 0.8132$ is a constant. The corresponding values of $\Gamma$ are also displayed (in right vertical axis) in Fig. S\ref{figS1}. It is noteworthy that for substrates expanding linearly in time ($\gamma=1$) the parameter $\Gamma$ (let us call it $\Gamma^*$) has a negligible dependence on $\omega$ and is the same as for flat (fixed-size - $\omega=0$) systems \cite{Ismael14}. Thus, for general $\gamma$, one may write
\begin{equation}
\Gamma(\gamma,\omega)=\Gamma^*+(\gamma-1) g(\gamma,\omega), 
\end{equation}
where $g(\gamma,\omega)$ is a positive function, so that $\Gamma < \Gamma^*$ ($\Gamma > \Gamma^*$) if $\gamma<1$ ($\gamma>1$), see Fig. S\ref{figS1}.

\section{Rescaling of the spatial covariance}
\label{secampcov}

Dynamic scaling predicts that $C_S/(\Gamma t)^{2\beta} \simeq \Phi(x)$, with $x \equiv A r^{2\alpha}/2 (\Gamma t)^{2\beta}$, where $A$ is the scaling amplitude of the height-height correlation function - defined as $C_H(r,t) \equiv \left\langle [h(x+r,t) - h(x,t)]^2 \right\rangle$ - with the distance $r$ [$C_H(r) \simeq A r^{2\alpha}$]. The scaling function $\Phi(x)$ is expected to be universal, being given by the Airy$_2$ (Airy$_1$) covariance in circular (flat) KPZ interfaces evolving in the plane. According to the KPZ theory, for 1D systems, $\Gamma = |\lambda| A^{2}/2$ \cite{Krug1992}, where $\lambda$ is the coefficient of the nonlinear term in the KPZ equation, usually referred to as the velocity excess. Thence, the ($\gamma,\omega$)-dependence in $\Gamma$ discussed above could imply in a similar dependence in $A$ and/or in $\lambda$. Unfortunately, it is hard to estimate accurately the value of $A$ from the scaling of $C_H$, which usually suffer from strong finite-size and -time corrections. We have indeed tried to do this, but the results (not shown) obtained do not allow us to conclude whether $A$ is a constant or ($\gamma,\omega$)-dependent. Furthermore, it is not clear to us how to determine the coefficient $\lambda$ in our expanding systems, because the tilting method \cite{barabasi} would not work here. Namely, by tilting the substrates, as they expand due to column duplications, their slope would decrease in time.

However, it does not matter what is the rescaling used, if it is appropriately applied to all data, it shall allow us to compare the covariances for different systems and to determine if they agree with Airy$_1$, Airy$_2$ or whatever. Thereby, in our work we compare the covariances by plotting $C_S/w_2$ against $r/a$, where $a$ is chosen so that at $r/a=1$ all curves have $C_S/w_2=0.2$. So, since $C_S(r=0)=w_2$, our rescaling obliges all curves to start at $C_S/w_2=1$ and to pass at $C_S/w_2=0.2$ at $r/a=1$. We remark that, if instead of the arbitrary $0.2$ point we had used $0.1$ or any other positive value, our conclusions on the covariances' behavior would still be the same.

\section{Derivation of the correction in the KPZ ansatz due to column duplication}
\label{seccorrec}

As demonstrated in Ref. \cite{Ismael14}, the KPZ non-linearity allied to the column duplication process yields a correction in the KPZ ansatz (Eq. \ref{eqansatz}), which is logarithmic when the substrate expands linearly in time. Here, we will generalize this by determining the main corrections for any $\gamma$. Let us start noticing that in systems with periodic boundary conditions the random column duplications cannot create a global curvature in the interface, so that $\langle \nabla^2 h \rangle=0$, but they can affect the nonlinear term $\langle (\nabla h)^2 \rangle$.

\begin{figure}[!t]
\includegraphics[width=7.cm]{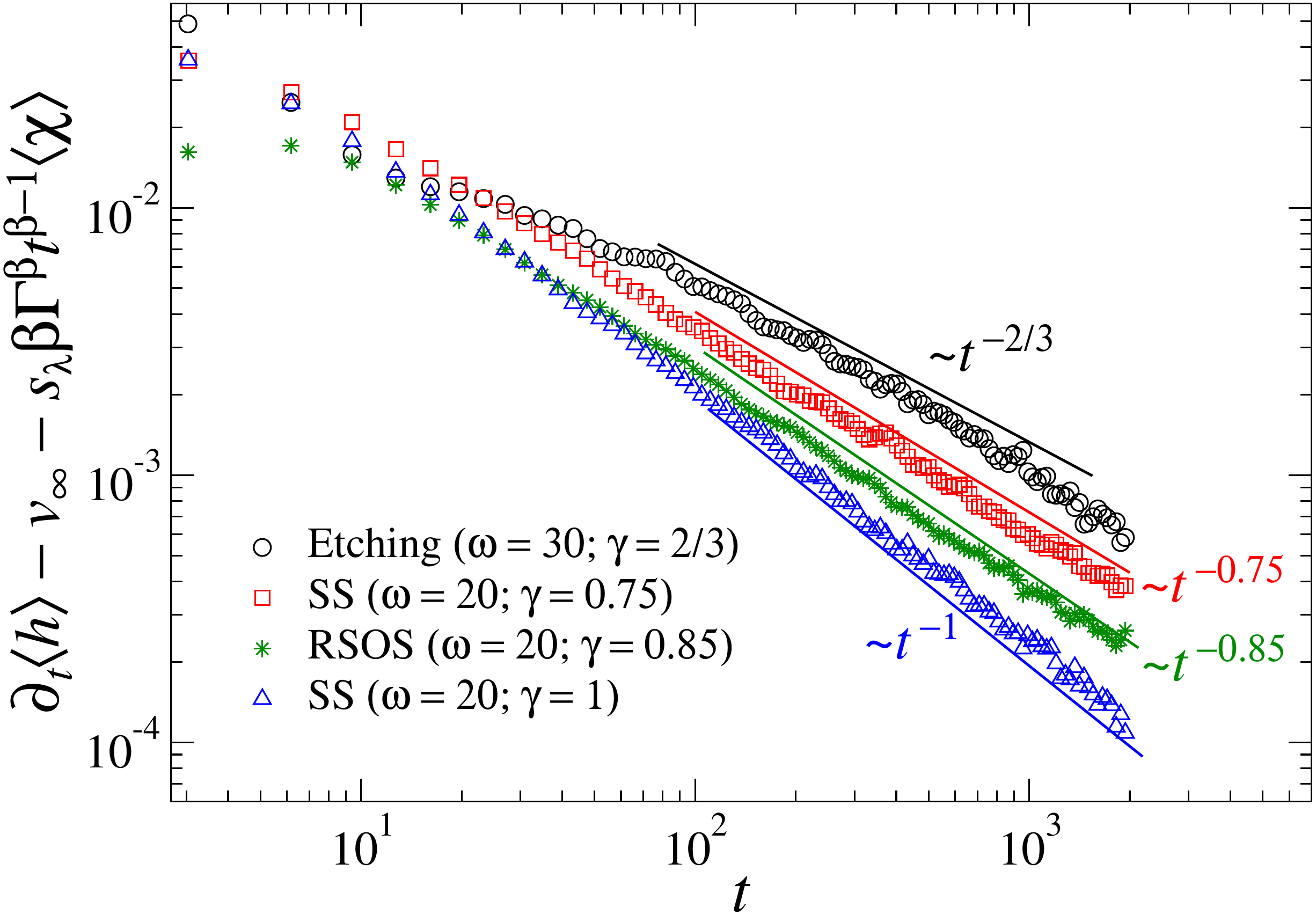}
\caption{(Color online) Temporal variation of the correction $Q \equiv \partial_t \langle h \rangle - v_{\infty} + \beta s_{\lambda}\Gamma^{\beta} t^{\beta-1} \langle \chi \rangle$ for several models and parameters. The lines have the indicated slopes.}
\label{figS2}
\end{figure}

On a $d_s$-dimensional substrate, the average of the local gradient at time $t$ is given by
\begin{equation}
G_t=\frac{1}{L^{d_s}}\sum_{i=1}^{L^{d_s}} (\nabla h_i)^2.
\label{eqGt}
\end{equation}
If the number of duplications occurring in a time unity is (on average) $l=\gamma \omega t^{\gamma-1}$, then, at time $t+1$, one has
\begin{equation}
G_{t+1}=\frac{1}{(L+l)^{d_s}}\left[ \sum_{i=1}^{L^{d_s}}(\nabla h_i)^2+\sum_{i=L^{d_s}+1}^{(L+l)^{d_s}}(\nabla h_i)^2 \right],
\label{eqGt+1}
\end{equation}
where the first (second) summation in the rhs runs over the non-duplicated (duplicated) columns in that time interval. Disregarding the effects of particle deposition and considering the statistical equivalence of sites, one finds
\begin{equation}
\sum_{i=L^{d_s}+1}^{(L+l)^{d_s}}(\nabla h_i)^2 \approx\frac{d_s-1}{d_s} \left[ (L+l)^{d_s}-L^{d_s}\right]G_t,
\label{eqGradm}
\end{equation}
where the ratio $(d_s-1)/d_s$ is due to the local smoothing caused by the duplications, since immediately after the duplication of a column (say $j$) in a given direction $x_k$ one has $\left( \partial h_j/\partial x_k\right) =0$. From Eqs. \ref{eqGt} and \ref{eqGradm}, one may re-write \ref{eqGt+1} as
\begin{equation}
G_{t+1}\approx\left( 1-\frac{1}{d_s}\left[ 1-\left(\frac{L}{L+l} \right)^{d_s} \right]\right)G_t.
\end{equation}
Now, considering long times, such that $L\approx \omega t^{\gamma}$ and so $\frac{L+l}{L} \approx 1+\frac{\gamma}{t}$, and disregarding terms $\mathcal{O}(t^{-2})$, one obtains
$G_{t+1}-G_t\approx-\frac{\gamma}{t}G_t \quad \textnormal{or}\quad \frac{d G}{d t} \approx-\frac{\gamma}{t}G$,
which leads to $G_t\sim t^{-\gamma}$. Finally, since $\frac{\partial h}{\partial t} \sim G_t \sim t^{-\gamma}$ in the KPZ equation, a correction term $\zeta t^{1-\gamma}$ is expected in the KPZ ansatz (Eq. \ref{eqansatz}), where $\zeta$ is in principle a stochastic variable. Note that for $\gamma=1$ it turns out to be the logarithmic found in \cite{Ismael14}.

Therefore, the average height of the interfaces are expected to evolve as
\begin{equation}
\langle h \rangle =v_{\infty}t+s_{\lambda}(\Gamma t)^{\beta} \langle \chi \rangle + \langle \eta \rangle + \langle \zeta \rangle t^{1-\gamma} + \ldots, 
\label{eqansatz}
\end{equation}
where we have included also the well-known additional constant correction $\eta$ \cite{Takeuchi2010,Sasamoto2010,Alves11,Ismael14}. From the derivative of this equation in time, we find that $Q \equiv \partial_t \langle h \rangle - v_{\infty} + \beta s_{\lambda}\Gamma^{\beta} t^{\beta-1} \langle \chi \rangle \simeq (1-\gamma) \langle \zeta \rangle t^{-\gamma}$. This is indeed confirmed in Fig. S\ref{figS2}, where one sees that $Q$ decays in time as $Q \sim t^{-\gamma}$.

\section{Spatial covariance in random depositions out of the plane}
\label{secRD}

\begin{figure}[!h]
\includegraphics[width=7.cm]{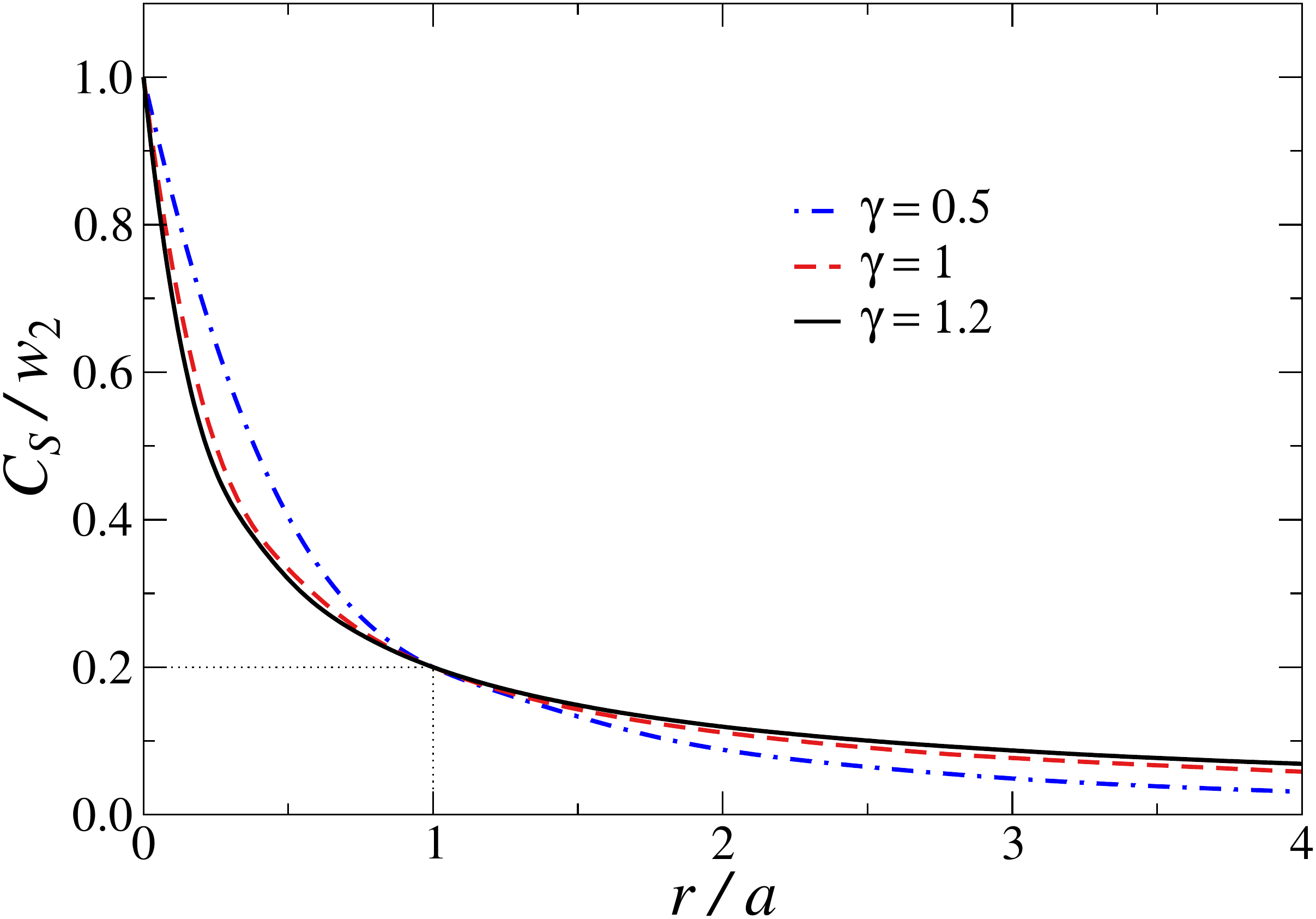}
\caption{(Color online) Rescaled spatial covariance for the random deposition model for the exponents $\gamma$ indicated.}
\label{figS3}
\end{figure}

In order to analyze the spatial correlations generated by column duplications, we have performed simulations of a random deposition process \cite{barabasi} on substrates expanding as $\left\langle L(t) \right\rangle = L_0 + \omega t^{\gamma}$, following the same method used for the other models. In the RD model the aggregation rule is always simply $h_i \rightarrow h_i + 1$, where $i$ is a randomly chosen site, so that no correlation is generated by the deposition process. Indeed, $C_S(r,t)=0$ for any $r>0$ for this model in the flat ($\omega=0$) case. For expanding interfaces, however, we find non-trivial correlations. Although the rescaled covariance curves have negligible dependence on $\omega$ and on time, they change with the exponent $\gamma$ (see Fig. S\ref{figS3}). The curve for $\gamma=1.2$ is compared with those for the other models in Fig. 4d of the main manuscript.

\end{document}